\documentclass[
 reprint,
 amsmath,amssymb,
aps,
]{revtex4-1}

\usepackage{caption}
\usepackage{subcaption}
\usepackage{amsmath}
\usepackage{graphicx}% Include figure files
\usepackage{dcolumn}% Align table columns on decimal point
\usepackage{bm}% bold math
\usepackage{float}
\usepackage{color}

\usepackage{hyperref}% add hypertext capabilities
\usepackage[mathlines]{lineno}% Enable numbering of text and display math
\begin{document}

%\preprint{APS/123-QED}

\title{Collimating cylindrical Surface Leaky Waves for Highly Improved Radiation Characteristics of Holograms }% Force line breaks with \\

\author{Mohammad Moein Moeini}
 \email{mohammad\_moeini@elec.iust.ac.ir}

 %\altaffiliation[Also at ]{}%Lines break automatically or can be forced with \\
\author{Homayoon Orazi}%
 \email{h\_oraizi@iust.ac.ir}
\author{Amrollah Amini}%
 \email{amini\_am@elec.iust.ac.ir}
\affiliation{%
 Department of Electrical Engineering, Iran University of Science and Technology, 
 \\1684613114, Tehran, Iran%\\
}%


%\affiliation{
%\affiliation{
 %Third institution, the second for Charlie Author
%\affiliation{%



\begin{abstract}
High resolution microwave leaky wave holograms excited by center-fed cylindrical surface wave launcher show a null at object wave direction which is an undesired effect for electromagnetic beamforming. Also,
planar leaky wave metasurfaces generating a tilted beam extremely suffer from the destructive effect of non-forward surface leaky waves at frequencies other than the design frequency and they are almost operable at a single frequency. Here we propose a 2-D modified hologram configuration using parabolic surface reflector to collimate the non-forward leaky modes into the forward leaky modes. The modified hologram presents null-free radiation pattern and highly improved operating frequency bandwidth. The consequent frequency bandwidth provides the scannabality property by frequency variation. The forward mode-dominant surface wave excitation of hologram lets the metasurface to generate the object beam more precisely; therefore, high directivity all over the operating bandwidth is obtained. The parabolic reflector lets the radiative surface of the metasurface to get shrunken in less than half of the conventional holograms. The concept is verified by fabrication and experimentally tested confirming the beam maintenance over a reasonable frequency range and scannability property. 
%\begin{description}
%\end{description}
\end{abstract}

\pacs{Valid PACS appear here}% PACS, the Physics and Astronomy
\maketitle

%\tableofcontents

\section{\label{sec:level1}Introduction}
Controlling electromagnetic wave and making it to radiate in a desired manner have been always attracting tremendous attention in applied electromagnetics; the advent of metamaterials and metasurfaces \cite{pendry,engheta} was a breakthrough in radiation engineering. Design and engineering of modulated metamaterials to control the scattered \cite{yu} or leaked \cite{oliner,sievenpiper} electromagnetic wave have been raising great attention in recent literatures \cite{galli,kevin,werner}. The electromagnetic wave can be manipulated through metasurfaces to obtain certain properties. Intoduction of leaky waves in electromagnetics has a great importance for beam engineering \cite{marcuvitz}. Power leakage engineering using metasurfaces based on variables dependent on the interaction of power generated by surface wave launcher and  the metasurface is an amazing solution to modern beamforming necessities like frequency-modulated continuous wave radars. Gabor firstly introduced the theory of holography in optics in 1948 \cite{gabor}. Holography was restricted to optical spectrum for many years until Chaccacci proposed the holographic principle for antenna design and microwave regime applicability of the theory in 1968 \cite{checcacci_68}. In a sinuously modulated surface where the modulation process governs by the holographic principle, the precise knowledge of the surface impedance on the holographic surface yields exact electromagnetic fields above the surface. In fact, the waveguide with a partially reflective interface made of periodic perturbations lets the horizontally guided wave to penetrate off the interface to form the object wave \cite{collin,itoh}. As the wave travels longer in waveguide, the object wave builded up from the gradual wave leakage approaches more accuracy. The concept of holographic leakage engineering for obtaining a desired object wave was proved by Sievenpiper \cite{sievenpiper}. On 2-D holographic metasurfaces which are designed to generate a tilted object wave, a cylindrical surface wave launcher can be selected to excite the modulated surface impedance. Metalic patches can realize the required surface impedance for holographic beamforming applying the modulation on patch dimensions. The hexagonal metallic patch unit-cell can be chosen as an isotropic unit-cell to realize the surface impedance.

For a center-fed leaky wave hologram,
while the surface impedance is realized with low resolution the object wave construction is possible. However when the holographic surface impedance is realized with high resolution the radiation pattern tends to approach zero value at the direction of object wave. This will be proved in next section.

The surface wave that has opposite horizontal wave number vector relative to the desired leaky wave is classified as the backward surface wave \cite{collin,itoh,monticone}. The wave leakage corresponding to the backward surface wave in a leaky wave system deviates from the leaked wave from forward surface wave when the operating frequency leaves the main design frequency \cite{cui,nannetti}. Therefore, the holographic leaky wave systems are capable of beamforming in a narrow vicinity around the design frequency \cite{minatti,minatti1}. Center-fed modulated surfaces have been always suffering from the single frequency operation  \cite{sievenpiper}. In \cite{cui}, side-fed hologram is tried to overcome the problem, however, induced current on the edges around the surface wave launcher causes destructive effect on the generated object wave. Rearranging  cylindrical surface leaky wave in a way that the forward surface wave participates more than the non-forward surface wave in object wave forming might be the most efficient way to provide wide-band holograms. In case that the leaky wave holograms are operable at frequencies except the preliminary design frequency, the main beam starts to scan the elevation angle \cite{moeini}. This scannabality provided by the reflector-enabled leaky wave metasurfaces, makes them a suitable choice for frequency dependent scanning systems.

In this work, we combine the concept of the non-forward surface wave redirection for ideal wave leakage and the holography theorem. We propose parabolic surface wave reflector for perfect non-forward mode suppression. The idea of a parabolic perfect electric conductor (PEC) reflecting boundary located next to the cylindrical  surface wave launcher, makes the redirection of 2-D bessel formed surface wave to almost 1-D forward surface wave possible. With this technique, all the energy generated by the cylindrical wave launcher is manipulated to participate in the beamforming process as the forward leaky mode. The theory of the holography is applicable to the new design only considering the modified surface wave on the metasurface with the presence of parabolic reflector. Therefore, the resultant all-forward surface wave, provides a wide operating frequency band over 13-18 GHz. The consequent scannabality of the elevation angle is obtained. The absence of non-forward modes in the surface wave exciting the holographic surface impedance, lets the wave leakage to focus on object wave generation more accurately than a conventional hologram. A gain above the 22dB is obtained for the reflector-enabled hologram over the operating frequency. The antenna radiation scans the elevation angle from $33^\circ$ to $69^\circ$.

\section{\label{sec:level2}Holography Principle}

\begin{figure}[H]
\centering
\includegraphics[width=0.9\linewidth]{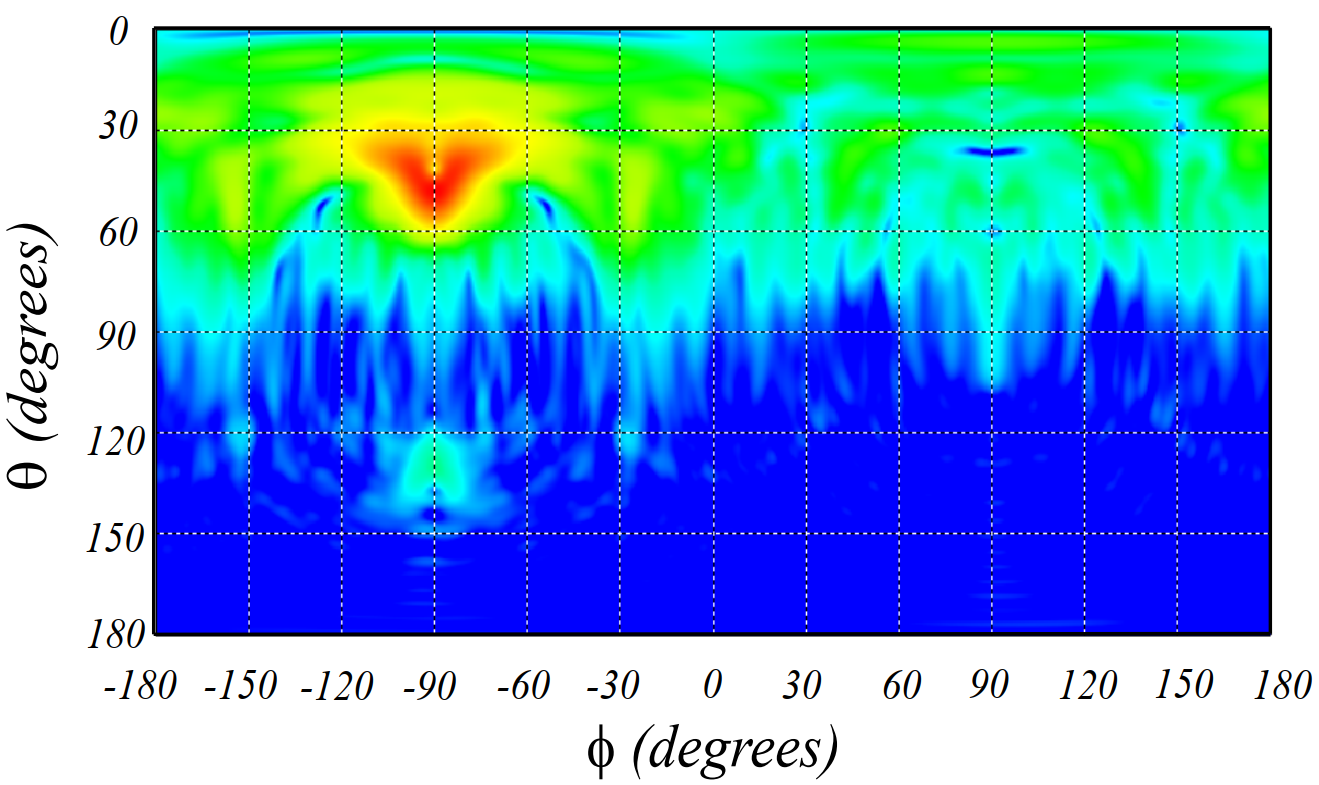}
\caption{The distributed 2-D radiation pattern of the high resolution conventional radiative hologram illustrating the undesired null in object wave direction at the far-zone electric field pattern.}
\label{fig:fig01}
\end{figure}

When a surface is illuminated by two electromagnetic waves, they may interact with each other differently depending on the capture point. The obtained interaction is due to the reinforcing or canceling at various positions on the surface \cite{hariharan1}. The holography principle is based on recording the intensity resulted from the interference process. The theory says by exciting the recorded interference using one of the waves participated in the interference production, the other one will be constructed. Reference wave $\psi_{ref}$ excites the recorded interference to construct object wave $\psi_{obj}$. The reference and object waves often stand for the input wave in forms of incident wave and the scattered wave from the modulated surface, respectively. However, in our application the reference wave is surface wave traveling on the modulated surface and the object wave is made of leakage through the path which the reference wave moves on. The resultant intensity of the two waves interaction is \cite{hariharan2}:
\begin{equation}
\label{eq:eq01}
I=|\psi_{ref}+\psi_{obj}|^2
\end{equation}
After some simplifications the intensity will be:
\begin{equation}
\label{eq:eq02}
I=|\psi_{ref}|^2+|\psi_{obj}|^2 + 2|\psi_{ref}||\psi_{obj}|\cos(\phi_{ref} - \phi_{obj})
\end{equation}
$\phi$ indicates the wave phase value. We will use the surface impedance parameter as the recorded interference of two interacting waves. However, there may be other options regarding to the wavelength. To represent the interference in form of the surface impedance, the (\ref{eq:eq02}) can be rewritten as:
\begin{equation}
 Z_s = jX_0(1+M\times\Re\{\psi_{ref}\psi_{obj}^*\})
 \label{eq3}
 \end{equation}
 Where $X_0$ and $M$ are average surface impedance and modulation factor respectively. A monopole surface wave launcher is suitable for exciting the surface impedance. In fact, the monopole radiates like a long conducting wire on the metasurface $(z=0)$. A long conducting wire carrying $I=I_0\hat{a_z}$ current, radiates $TM$ wave propagating in radial direction. For metasurface applications we often are interested in $TM_0$ mode. Electric and magnetic fields produced by a cylindrical surface wave launcher for an isotropic surface impedance is:
 \begin{equation} \label{eq01}
\textbf{H}_{z=0^+}=J{sw}H_1^{(2)}(k_{sw}\rho)(-\hat{\phi})
\end{equation}
\begin{equation} \label{eq05}
\textbf{E}_{z=0^+}=Z_sJ{sw}H_1^{(2)}(k_{sw}\rho)\hat{\rho}
\end{equation}
Therefore, the reference wave form can be treated as $H_1^{(2)}(k_{sw}\rho)$. The asymptotic evaluation of the reference wave is:
\begin{equation}
\label{asymptotic}
H_1^{(2)}(k_{sw}\rho)=a_0\sqrt{\frac{2}{\pi \sqrt{\beta_{sw}^2+\alpha_{sw}^2}\rho}}e^{-j\beta_{sw}\rho}e^{-\alpha_{sw}\rho}
\end{equation}
Where $k_{sw}=\beta_{sw}-j\alpha_{sw}$ and $a_0$ is a complex constant value. The surface impedance in (\ref{eq3}) requires the interactive waves to be normalized. And finally the reference wave can be written as  \cite{sievenpiper}:
\begin{equation}
\label{eq:eq05}
\psi_{ref}=e^{-j\beta_{sw}\rho}
\end{equation}
 $\beta_{sw}$ can be approximated for small values of $M$ as below \cite{patel}:
\begin{equation}
\label{eq:eq05}
\Re\{k_{sw}\}=\beta_{sw} = k_0 \sqrt{1+X_0^2}
\end{equation}
We are seeking for generating an object wave which propagates in $\theta_0$ and $\phi_0$ spherical angles direction. The object wave expression in the desired direction is defined as:
\begin{equation}
\label{eq:eq08}
\psi_{obj}=e^{-j(kx\sin\theta_0 \cos\phi_0 + ky\sin\theta_0 \sin\phi_0)}
\end{equation}
The surface impedance pattern in this case is:
\begin{equation}
\label{eq:eq06}
\begin{split}
Z_s=jX_0(1+M\sin(\beta_{sw}\rho-k_0 x \sin\theta_0\cos\phi_0-\\
k_0 y \sin\theta_0\sin\phi_0)-\psi_0+\frac{\pi}{2})
\end{split}
\end{equation}
$\psi_0$ is the phase of $a_0$ in (\ref{asymptotic}). Electric aperture field just above the interactive surface impedance is \cite{minatti1}:
\begin{equation}
\label{eq:eq08}
\begin{split}
\textbf E_{apr}=\hat{\rho}E_{\rho 0}e^{-j(kx\sin\theta_0 \cos\phi_0 + ky\sin\theta_0 \sin\phi_0)}e^{-j(\psi_0-\frac{\pi}{2})}\times \\
\sqrt{\frac{2}{\pi \sqrt{\beta_{sw}^2+\alpha_{sw}^2}\rho'}}e^{-j\beta_{sw}\rho'}e^{-\alpha_{sw}\rho'}
\end{split}
\end{equation}
It should be noted that $E_{\rho 0}$ amplitude and $\psi_0$ phase are specified by the source parameters.
The primed letters are used to avoid misconception in the following analysis. The electric aperture field is obtained in a way that the attenuation constant $\alpha_{sw}$ represents the wave leakage. For a radiative aperture, the far-zone electric field is \cite{elliott}:
\begin{equation} \label{eq14}
\textbf E_{far}(r,\theta,\phi)\approx \frac{jk}{2\pi r}e^{-jkr}(f_\theta \hat{\theta}+f_{\phi}\hat{\phi)}
\end{equation}

\begin{gather*} 
f_\theta(\theta,\phi)=f_x\cos\phi+f_y \sin\phi ,\\ 
f_\phi(\theta,\phi)=\cos\theta (-f_x\sin\phi+f_y \cos\phi)
\end{gather*}
\begin{equation} \label{eq15}
\begin{split}
f_x=\int_{0}^{a}\int_{0}^{2\pi} (\textbf{E}_{apr}.\hat{x}) e^{jk(X_0\sin\theta\cos\phi+y'\sin\theta\sin\phi)}\rho'd\phi' d\rho'
\end{split}
\end{equation}

\begin{equation} \label{eq16}
\begin{split}
f_y=\int_{0}^{a}\int_{0}^{2\pi} (\textbf{E}_{apr}.{\hat{y}})e^{jk(X_0\sin\theta\cos\phi+y'\sin\theta\sin\phi)}\rho'd\phi' d\rho'
\end{split}
\end{equation}
Here we explore the electric field at the object wave direction $\theta_0$ and $\phi_0$. To fulfill this aim, derivations of $f_\theta(\theta_0,\phi_0)$ and $f_\phi(\theta_0,\phi_0)$ are necessary. Substituting $\textbf E_{apr}$ in $(13)$ and $(14)$ results:
\begin{equation} \label{eq15}
\begin{split}
f_\theta(\theta_0,\phi_0)=\int_{0}^{a}\int_{0}^{2\pi} E_{\rho 0}e^{-j\psi_0}\sqrt{\frac{2}{\pi \sqrt{\beta_{sw}^2+\alpha_{sw}^2}\rho'}} \\
\times e^{-\alpha_{sw}\rho'}\cos (\phi'-\phi_0)\rho'd\phi' d\rho'
\end{split}
\end{equation}
\begin{equation} \label{eq16}
\begin{split}
f_\phi(\theta_0,\phi_0)=\int_{0}^{a}\int_{0}^{2\pi} E_{\rho 0}e^{-j\psi_0}\sqrt{\frac{2}{\pi \sqrt{\beta_{sw}^2+\alpha_{sw}^2}\rho'}} \\
\times e^{-\alpha_{sw}\rho'}\sin (\phi'-\phi_0)\rho'd\phi' d\rho'
\end{split}
\end{equation}
The above integrations simply lead to zero meaning that the electric field vanishes at $\theta_0$ and $\phi_0$ which was the direction expected to  propagate the object wave.   Fig.\ref{fig:fig01} shows the distributed  pattern for a symmetric conventional center-fed hologram.
 The antenna is designed to radiate at $\theta_0=45^\circ$ and $\phi_0=-90^\circ$ at 18 GHz. As it is obvious, the holography theorem is unable to predict the null at the disign direction for a high resolution realized hologram at design frequency. The structure is designed based on the procedure intoduced in \cite{sievenpiper} using hexagonal unit-cells, which is even more isotropic compared to the rectangular patch \cite{Li_sievenpiper}, to realize the surface impedance. To achieve proper accuracy the side length of the hexagonal unit-cell is set to $1.7mm$. The integrations (\ref{eq15}) and (\ref{eq16}) are valid for a practical structure in case that the descretization resolution is high enough, otherwise the null at the design direction may disappear\citep{moeini}. Also, for holograms which avoid utilizing center-fed surface wave launcher, the null disappears due the disruption in surface wave distribution symmetry. Another design to remove the null is to form compensating phase correction by enforcing manual phase discontinuity on the surface impedance distribution \cite{Ettorre,Pandi}. All the suggestions offered to improve hologram operation are only at a narrow frequency band in the vicinity of the design frequency.
 
\begin{figure}[H]
\centering
\includegraphics[width=0.9\linewidth]{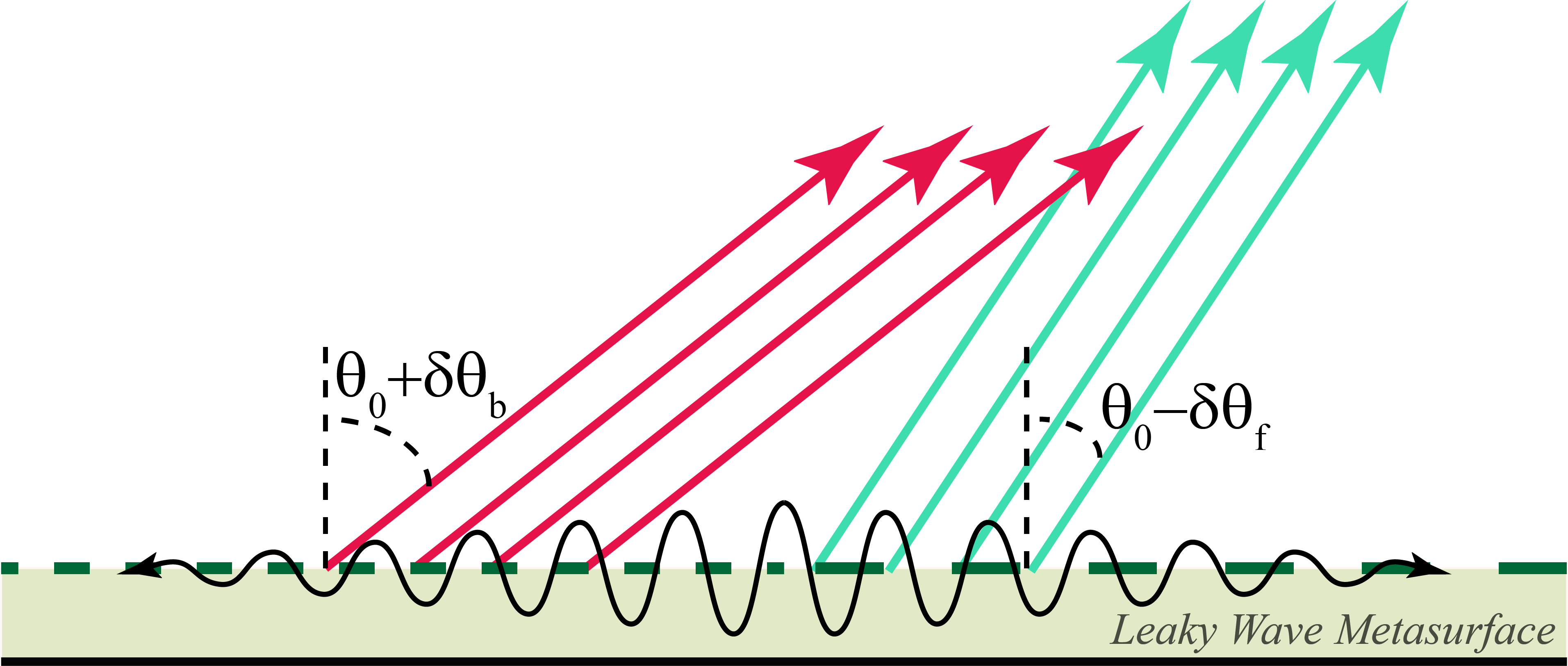}
\caption{Comparison of leaky waves from forward and backward surface waves on a 1-D metasurface. }
\label{1-D_forward&backward}
\end{figure}

\begin{figure}[H]
\centering
\includegraphics[width=1\linewidth]{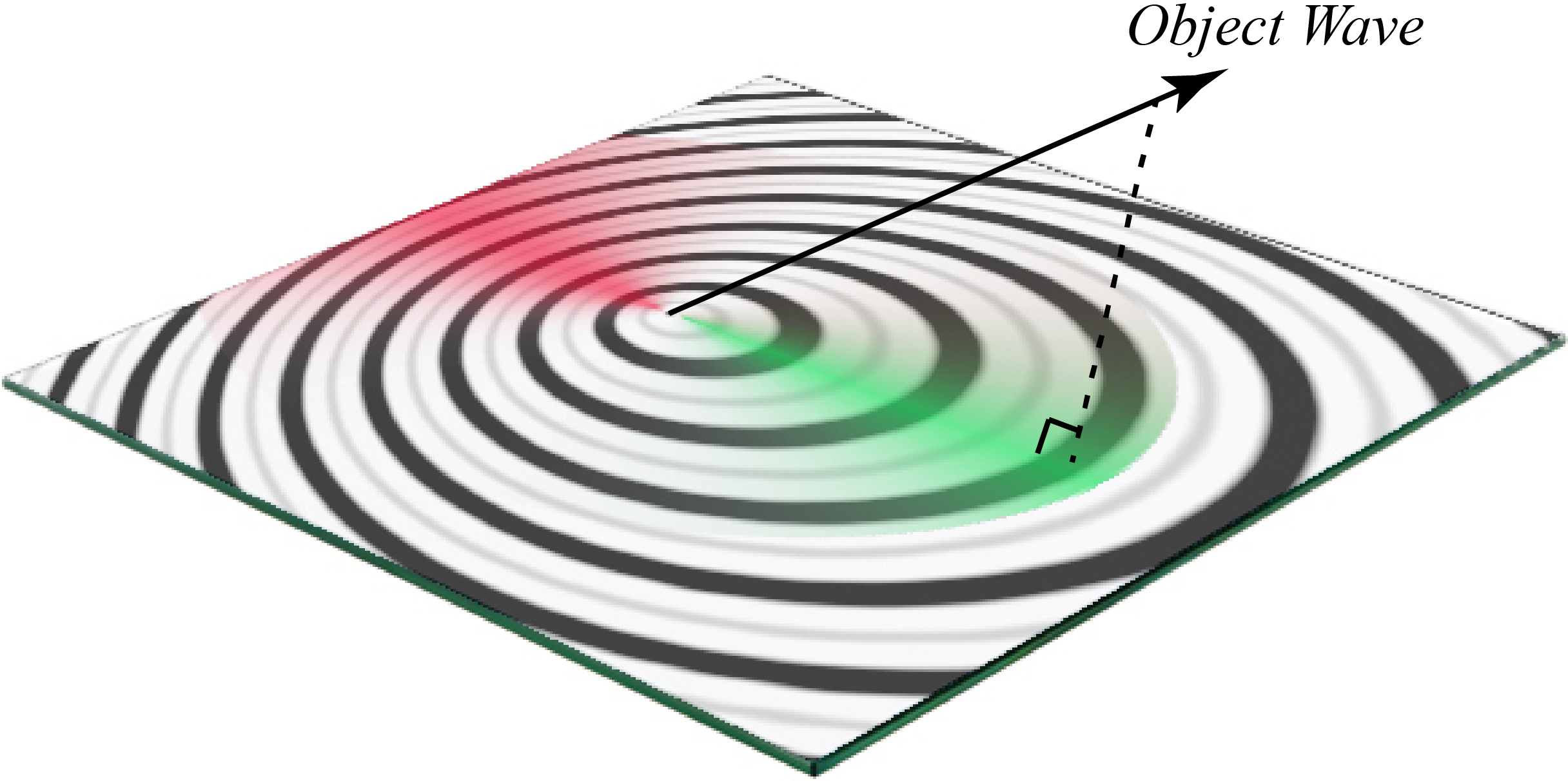}
\caption{Forward and backward surface leaky waves for a 2-D hologram supporting cylindrical surface wave excitation}
\label{green_red1}
\end{figure}

 The holographic leaky wave radiators are extremely restricted by frequency variation. Holographic antennas are often designed to generate object wave at a single frequency. However, a practical advantage of leaky wave radiators is the scannability over a frequency range. A physical interpretation of the limited operation frequency of holograms is a destructive effect called Rabbit's ears phenomenon \citep{cui}. On a 1-D x-directed leaky wave radiator fed using a surface wave launcher producing $\psi_{ref}=e^{-j\beta|x|}$ with sinuously modulation aimed to radiate in a tilted direction $\theta_0$, the periodicity of the surface impedance is different on forward and backward regions of the surface impedance. While the forward region is modulated to let the surface wave leaks at $\theta_0$, the backward region should be designed for beamforming at $90^{\circ}+\theta_0$. Then the surface impedance on the backward region needs to be modulated with lower periodicity than the forward region. The situation is ideal for operation at design frequency $f_0$. However, the wave leakage directions for each of the forward-region generated and backward-region generated waves show different propagation directions when the surface wave launcher starts to operate at a frequency slightly lower than $f_0$. For a launcher operating at $f_0-\delta f$, the forward and backward leaky waves propagate at $\theta_0-\delta\theta_f$ and $\theta_0+\delta\theta_b$, respectively (as illustrated in Fig.\ref{1-D_forward&backward}). By altering the frequency to other values except $f_0$, the surface wave fronts on two sides of the wave launcher face surface impedance steps forcing the surface waves on each side to leak in different directions. The resulted beams deviated from each other are like rabbit's ears and the phenomenon is called after that \cite{cui}.
This destructive effect of non-forward leaky waves at frequencies except design frequency on total radiated object wave can be generalized to 2-D leaky wave metasurfaces.
 Metasurfaces using monopole wave launcher radiate cylindrical surface wave fronts propagating along $\rho$ direction. Considering Fig.\ref{green_red1}, For a holographic metasurface aimed to produce a tilted object wave, project a picture of the object wave propagation path on the metasurface.  The projected picture on the metasurface is the path which ideal forward surface leaky wave is travelling on the waveguiding metasurface; showed in green on the metasurface in Fig.\ref{green_red1}. Moving on azimuth direction apart from the forward surface leaky wave, the destructive effect of the Rabbit's ears phenomenon increases until the worst case occurs for all-backward surface leaky wave path showed in red.
Therefore, any non-forward surface leaky wave cause destructive effect on bandwidth of the radiator. The conventional holograms use cylindrical wave launcher and leaky wave metasurfaces in microwaves and the presence of non-forward surface leaky waves is the reason of their highly narrow operating frequency.

\begin{figure}[H]
\includegraphics[width=0.9\linewidth]{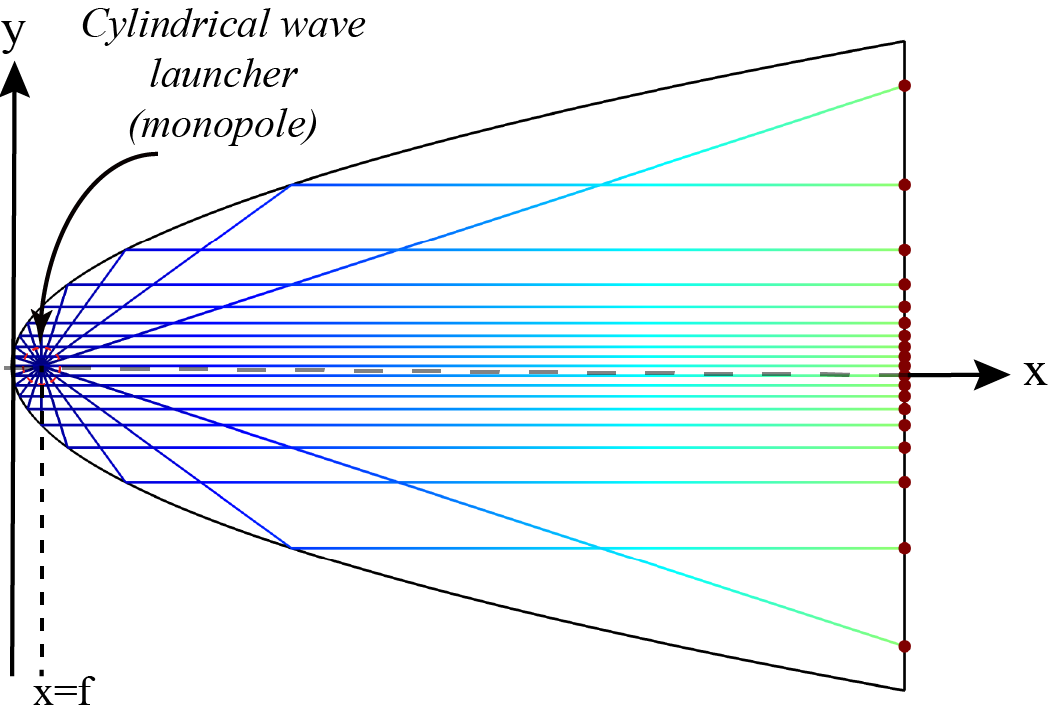}
\caption{Illustration of the scattered rays leaving a monopole radiator located at the focal point of parabolic PEC boundary.}
\label{fig:ray_tracing}
\end{figure}

 A parabolic reflector illuminated by a cylindrical wave launcher placed at the focal point of the parabola is a geometry that collimates incoming waves to parallel waves propagating in a single direction. The concept can be utilized to overcome the or Rabbit's ears phenomenon in leaky wave holograms. Referring to Fig.\ref{fig:ray_tracing}, half of the radiated waves in $x<f$ region are all redirected in forward surface leaky wave direction and rest of the leaving waves from monopole radiator in $x>f$ region are also changing their propagation path as they get reflected from the parabolic reflector. Thus, almost all of the surface leaky waves are collimated into the ideal forward surface leaky waves. It is important to notice that the power of the non-forward surface leaky waves is not eliminated; we are converting what was thought to be destructive for radiation characteristics of radiative holograms into ideal form of surface leaky waves.
\section{\label{sec:level3}Metasurface Design}
We introduce a reflector-enabled holographic metasurface. Placing a reflector next to the cylindrical surface wave launcher can break the symmetry of the surface wave configuration and results a null-free pattern at the design frequency $f_0$. On the other hand, by choosing the reflector profile as a parabolic surface wave reflector with a cylindrical surface wave launcher at the focal point $(f)$ of the parabola, the whole metasurface can be treated as a bundle of 1-D of leaky wave radiators which provide considerable  scannability property due to the perfect redirecting of almost all non-forward surface waves. Therefore, highly improved beam-width and directivity over a wide frequency range are predicted in comparison with conventional holographic antennas.
The preliminary holographic reference wave can be assumed (\ref{eq05}) when the parabolic reflector is not located on the structure. After embedding the parabolic reflector next to the surface wave launcher, the holographic reference wave is modified. The parabolic PEC boundary follows the equation:
\begin{equation}
\label{eq:eq06}
x=\dfrac{1}{4f}y^2
\end{equation}
Fig.\ref{fig:ray_tracing} provides an insight into the holographic reference wave configuration after the placement of the parabolic reflector on the hologram using ray optics. The cylindrical rays leaving the monopole surface wave launcher are collimated into a x-directed bundle of rays \cite{elliott}. These secondary reflected rays are all in-phase after passing the $x= f$ vertical line. Therefore, ignoring the wave function in small area between $x=0$ and $x=f$ lines, the reference wave at the other areas of the 2-D hologram can be assumed to be planary in-phase surface waves. For the sake of higher accuracy in reference wave estimation, the simulation results  are used to form the surface impedance using COMSOL Multiphysics \cite{comsol}.

\begin{figure}[H]
\includegraphics[width=1\linewidth]{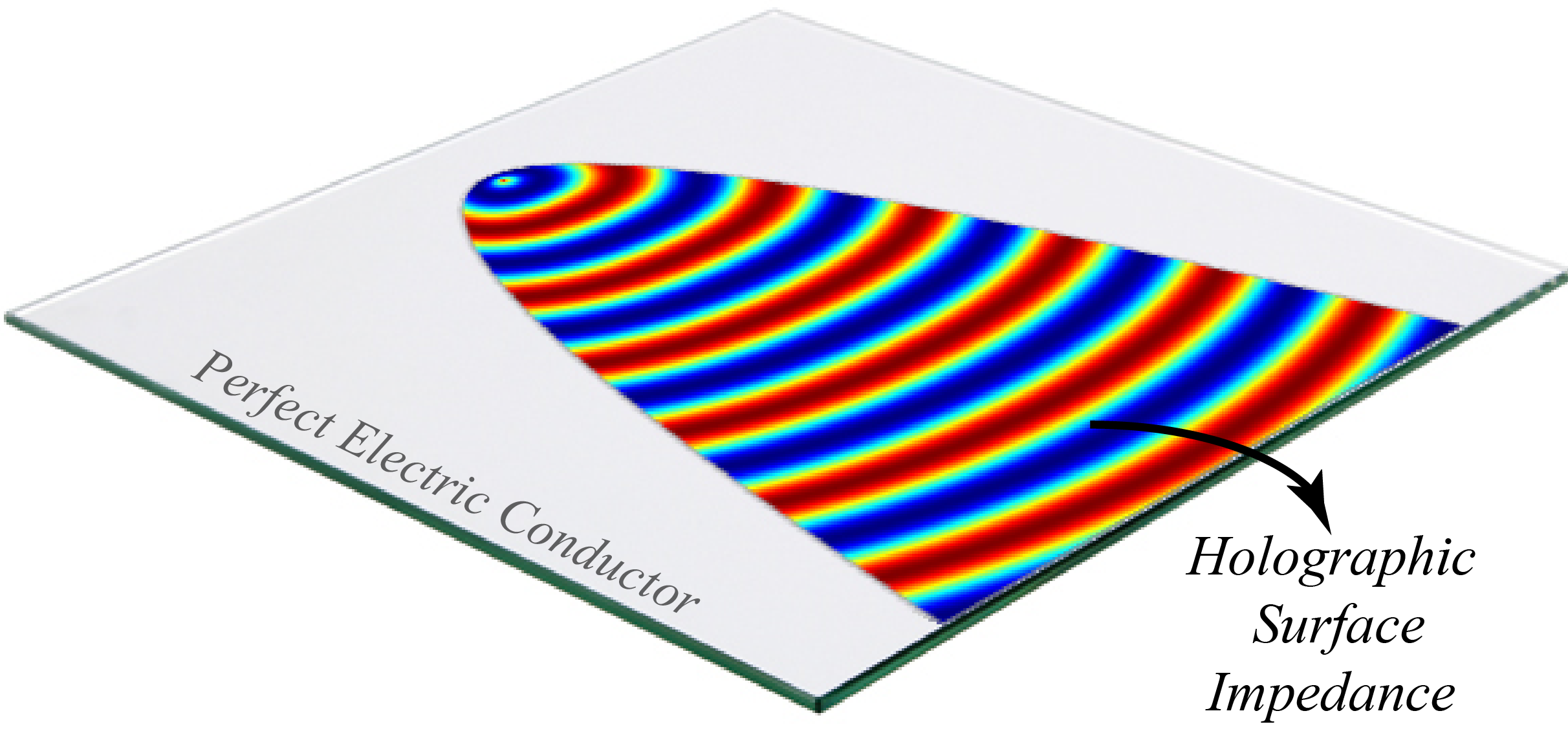}
\caption{Schematic 2-D distribution of the surface impedance on a parabolic reflector-enabled hologram.}
\label{vanishing_point}
\end{figure}

\begin{figure}[H]
\includegraphics[width=1\linewidth]{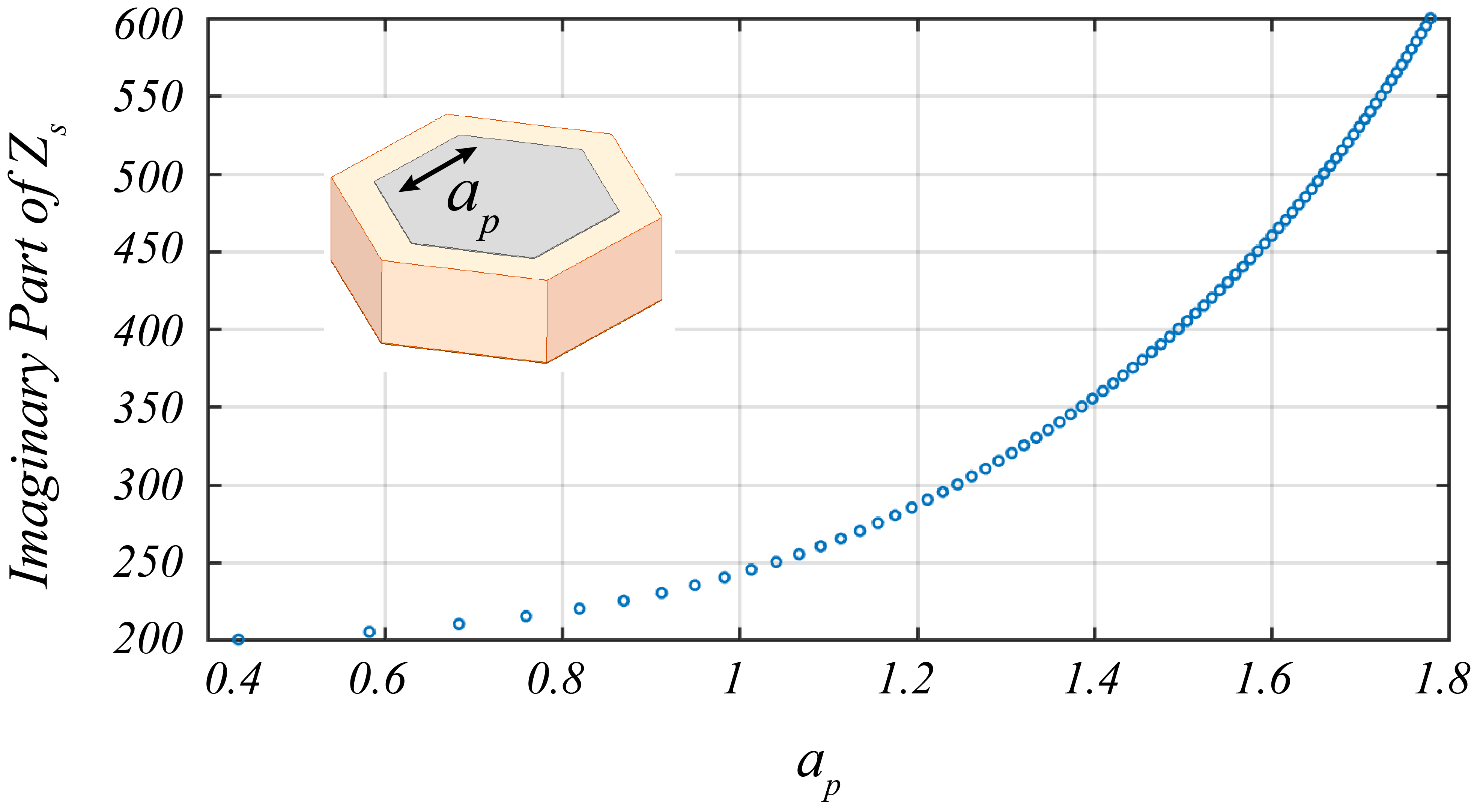}
\caption{The surface impedance imaginary value as a function of patch side length $(a_p)$.}
\label{realizing_curve}
\end{figure}

Knowing the reference and object waves leads to 2-D distribution of the surface impedance $(Z_s)$ using equation (\ref{eq3}). Once the surface impedance is derived, the theory for wave reconstruction is done. The resulted 2-D surface impedance distribution in presence of parabolic reflector is shown in Fig.\ref{vanishing_point}.
The theoretical holographic surface impedance needs to be realized to be practically capable of beamforming. We choose the hexagonal patch printed on the dielectric slab for the unit-cell, playing the role of surface impedance. The patch size needed for realizing the surface impedance is shown in Fig.\ref{realizing_curve}. For our application the total hexagonal unit-cell side length is chosen $1.7mm$ and the dielectric substrate electric permittivity is 3.55 with a thickness of $1.524mm$.

The modulation factor $M$  in (\ref{eq3}) determines the depth of sinuously space-dependent variations around the average surface impedance and it controls the attenuation constant $\alpha$. By defining M on low values, the surface wave obtains enough travel length on the surface and gradual wave leaking with proper accuracy. In contrast, choosing high values for M and consequent larger sinuously variations on the modulated surface make the surface wave leave the surface in a short amount of length that the surface wave travels on. Therefore, the leaking procedure does not acquire enough space and time to form the object wave with desired accuracy. High side-lobe levels are observed in this case. However, choosing very small values for M reduces the attenuation constant along the traveling path and object wave construction requires long traveling path. Therefore, there is a trade off between the object wave accuracy and practical waveguiding path length. 
We chose $M=0.35$ for our application. Since the parabolic reflector-enabled leaky wave metasurface collimates the surface wave, the surface leaky wave can be treated as array of unidirectional forward surface leaky waves. The radiation directivity for such a structure is\cite{collin}: 
\begin{equation}
\label{eq:eq12}
D\propto \frac{\beta_{\bot (-1)}}{\alpha_{\parallel (-1)}}=\frac{\sqrt{\omega^2\mu_0\epsilon_0-(\kappa+\frac{2\pi(-1)}{d})^2}}{\alpha_{\parallel (-1)}} 
\end{equation}
Where $\kappa$ is the fundamental harmonic of parallel propagation parameters including fundamental phase constant $\beta_{\parallel (0)}$ and  attenuation constant $\alpha_{\parallel (0)}$ $(\kappa=\beta_{\parallel (0)}-j\alpha_{\parallel (0)})$. $d$ is a period of the modulated surface impedance. $\beta_{\bot (-1)}$ is the phase constant of the leaked wave  and $\alpha_{\parallel (-1)}$ is the attenuation constant of wave in waveguiding metasurface.

 As the modulation factor $M$ increases, the surface wave tends to radiate sooner while traveling on the metasurface and the attenuation constant $\alpha_{\parallel (-1)}$ increases, which causes lower directivity value and object wave accuracy. Therefore, higher side-lobe levels are expected while the antenna scans the elevation angle.

\begin{figure}[H]
\includegraphics[width=1\linewidth]{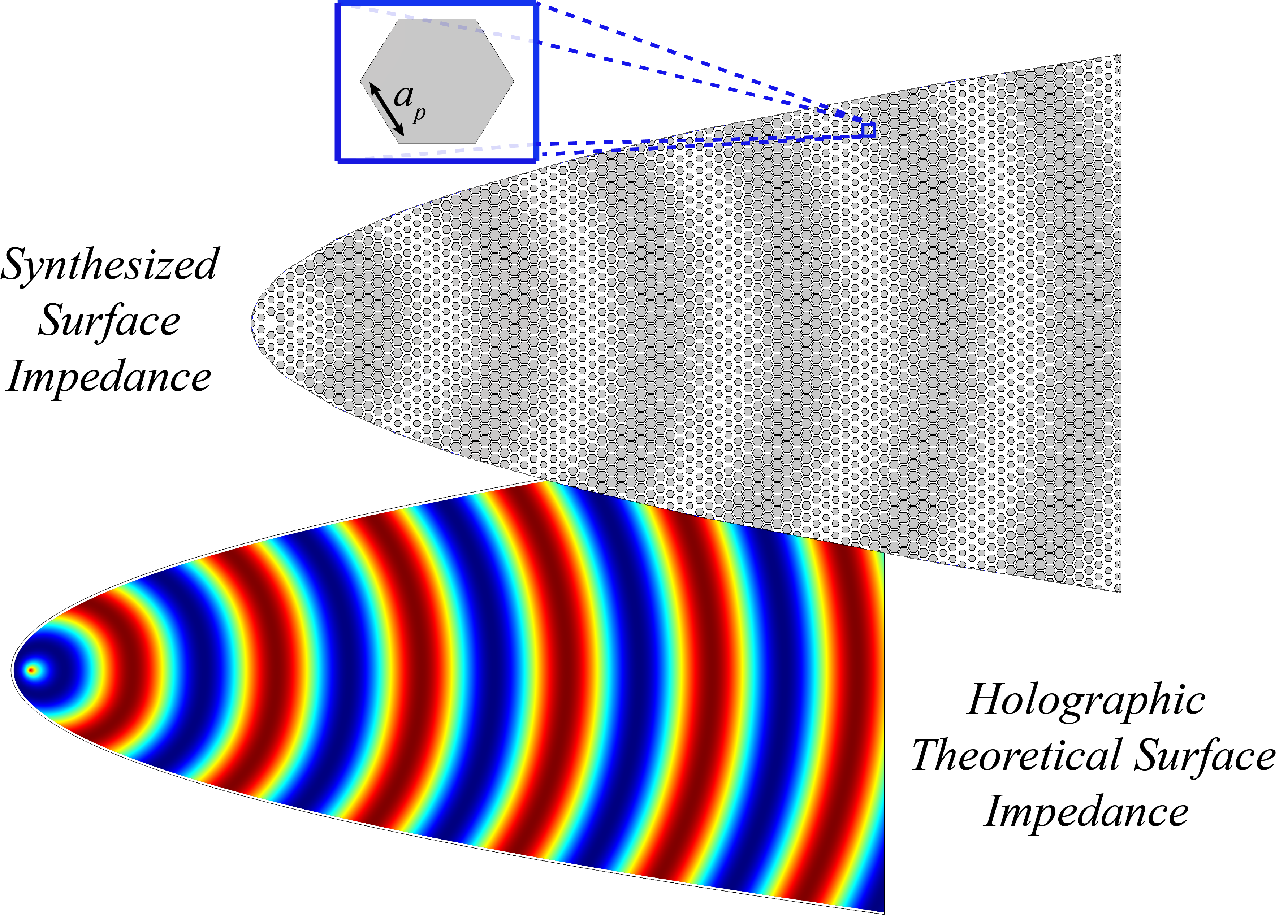}
\caption{Fully synthesized surface impedance for the holographic antenna with a parabolic surface wave reflector.}
\label{synthesized_impedance_comparision}
\end{figure}

\begin{figure}[H]
\centering
\includegraphics[width=1\linewidth]{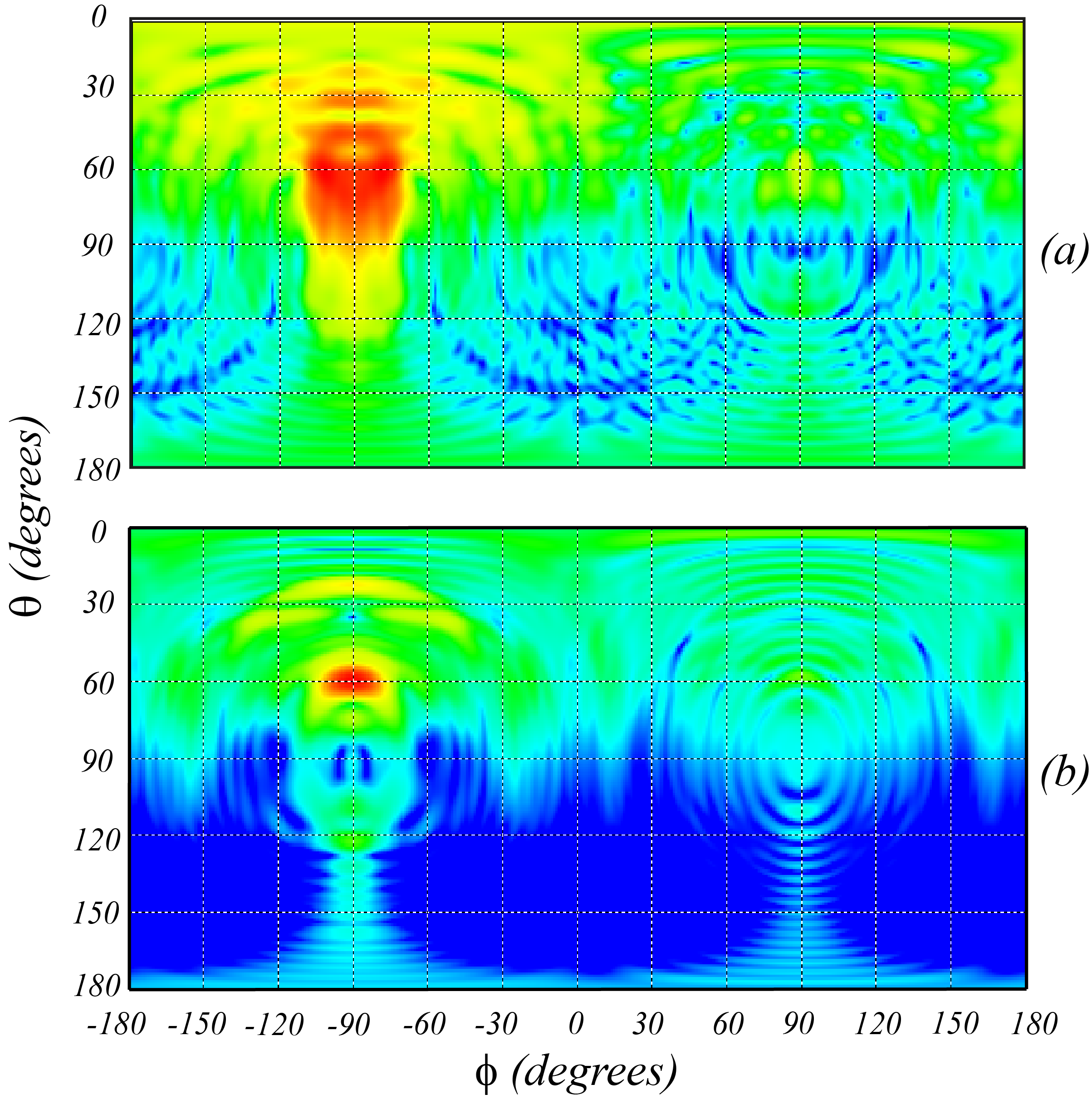}
\caption{Comparison of simulated \cite{CST} 2-D patterns of conventional$(a)$ and parabolic reflector-enabled$(b)$ holographic radiators at $f_0 -\delta f(17GHz)$.}
\label{pattern_comparision_17GHz}
\end{figure}

\begin{figure}[H]
\centering
\includegraphics[width=0.9\linewidth]{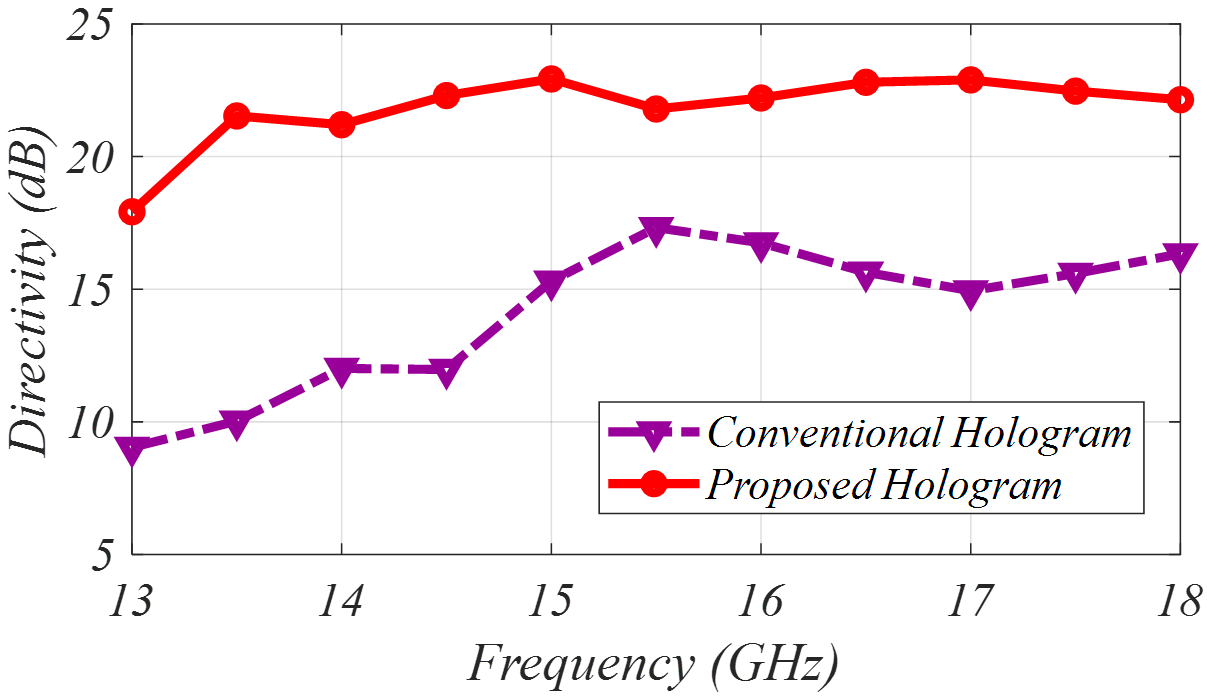}
\caption{Full-wave simulation of the conventional and proposed holograms with respect to frequency.}
\label{directivity_frequency}
\end{figure}

 Discretizing the 2-D surface impedance into subdomains each assigned to a certain surface impedance value is needed for practical realization. The hexagonal patches with different sizes show different value of average surface impedance. Therefore, the whole discretized 2-D surface impedance distribution can be synthetized using the corresponding hexagonal unit-cell for each subdomain. Fig.\ref{synthesized_impedance_comparision}, shows the fully synthesized model of the surface impedance.
To illustrate the concept some radiation characteristics of the conventional and proposed parabolic reflector-enabled holographic metasurface radiator both fed by cylindrical surface wave launchers are compared. The object wave direction and design frequency are assumed to be $\theta_0=70^{\circ}$ and $f_0=18GHz$  for both holograms. This selection of $\theta_0$ provides scanning property by frequency decrease.
To compare the effect of non-forward surface leaky waves on hologram operating bandwidth, the distributed 2-D radiation patterns at $f=17GHz$ are posed in Fig.\ref{pattern_comparision_17GHz}. Both holograms use hexagonal unit-cells to realize the surface impedance.
As it is clear in Fig.\ref{pattern_comparision_17GHz}, the non-forward surface leaky waves significant destructive effect on the object wave makes the radiation pattern unacceptable for the conventional hologram. In contrast, the hologram using parabolic reflector by redirecting the non-forward surface leaky waves into forward ones achieves a main-lobe magnitude of $22.9dB$ at $f=17GHz$ while the conventional hologram reaches $15 dB$. Fig.\ref{directivity_frequency}  illustrates  how the proposed hologram maintains highly directive over $13-18GHz$.

By lowering the operating frequency, the hologram tends to scan $\theta$ angle and $\theta_{scan}=60.0^{\circ}$ is obtained for $f=17GHz$.

\section{\label{sec:level4}Experimental Results}
\begin{figure}[H]
\centering
\includegraphics[width=.75\linewidth]{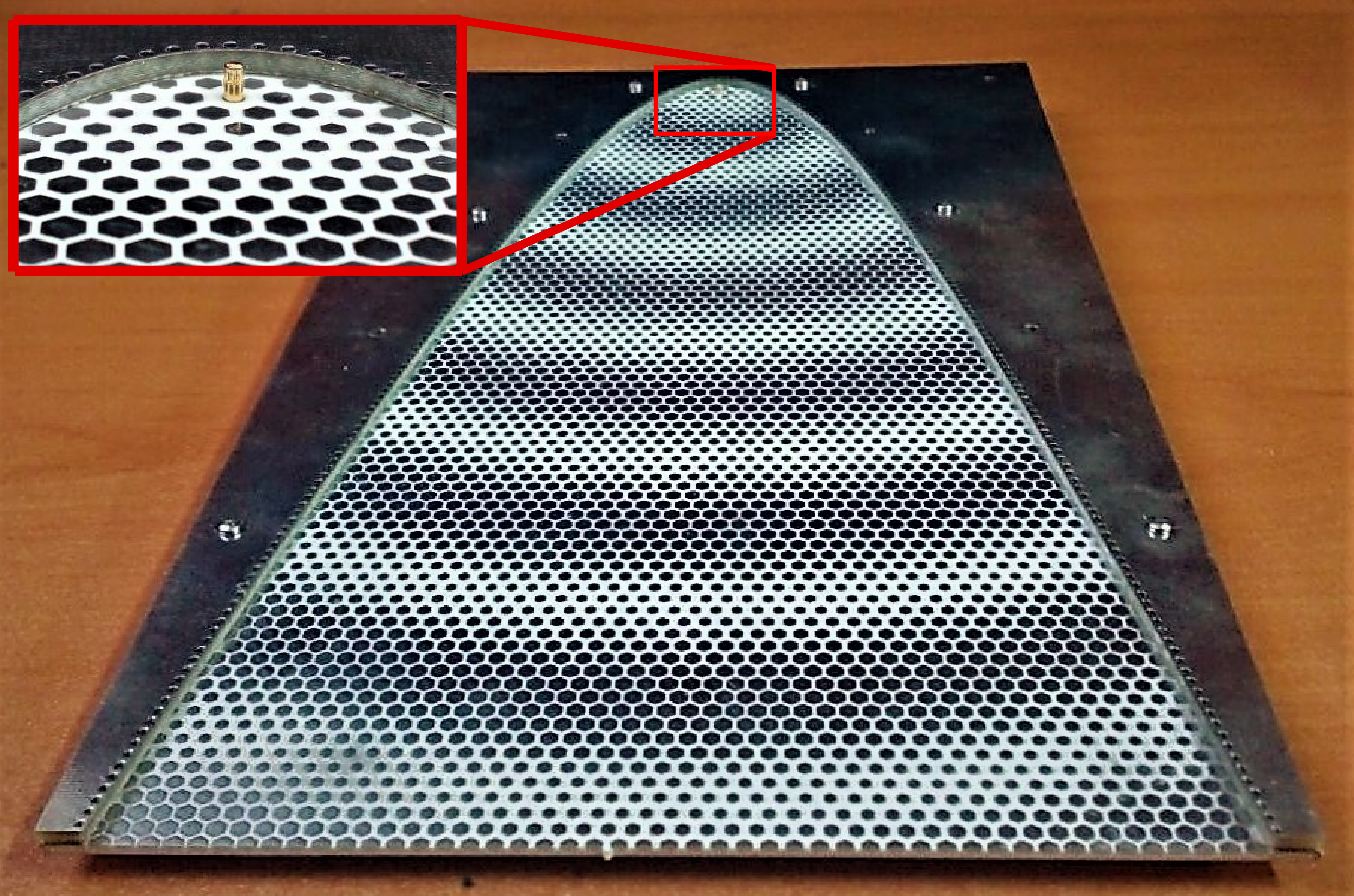}
\caption{Experimental setup of the proposed holographic metasurface aimed to generate a tilted object wave.}
\label{AX}
\end{figure}

\begin{figure}[H]
\includegraphics[width=1\linewidth]{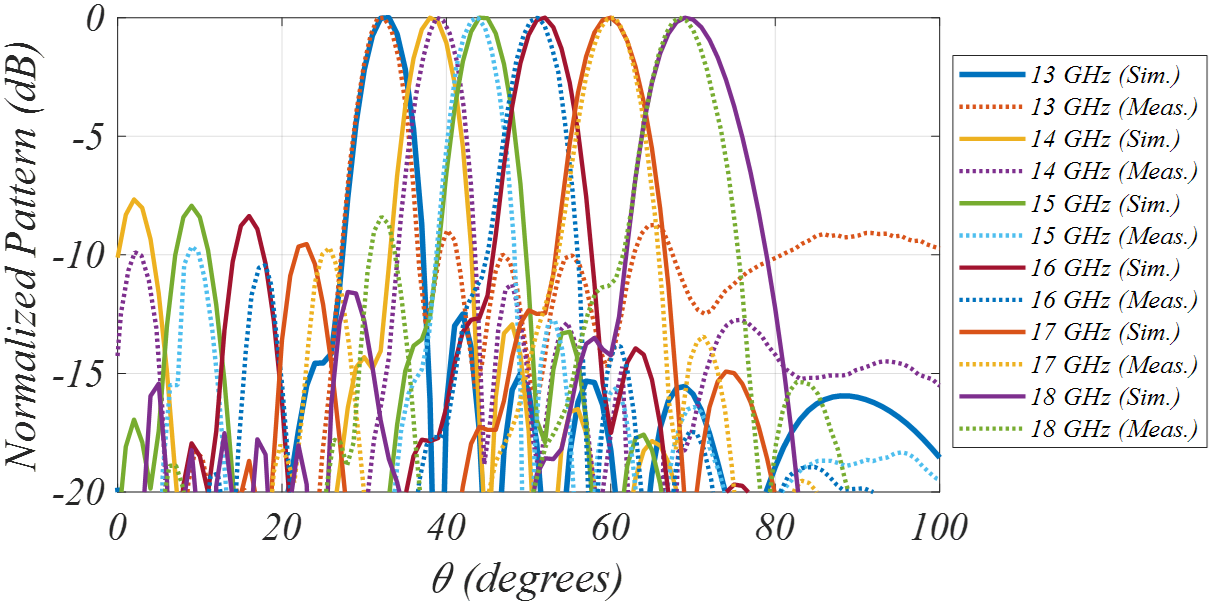}
\caption{Simulation(solid lines) and measurement(dashed lines) results for normalized directivity of the parabolic reflector-enabled holographic antenna.}
\label{Results}
\end{figure}

A holographic surface impedance calculated for object wave generation at $\theta_0$ scans the elevation angles by frequency variation. After surface impedance realization, the surface impedance 2-D distribution remains unchanged for all reference wave frequencies. When the monopole radiator frequency is set to a lower value than $f_0$, the realized surface impedance is then corresponding to an object wave propagating along a direction nearer to the normal axis to the metasurface plane (lower value of elevation angle $\theta_0$). The story is also true when the monopole radiator frequency is set at larger frequencies than $f_0$ and the elevation angle will be at a larger angle than $\theta_0$. Therefore,  a major facility provided by wide-band holographic metasurfaces is elevation angle scanning. 
The  wide-band holographic metasurface with the parabolic reflector needs to be realized using hexagonal unit-cells.
 Fig.\ref{AX} represents the fabricated prototype of the parabolic reflector-enabled hologram. The structure includes two layers, the waveguiding metasurface and the surface wave reflector. The metasurface is realized using hexagonal unit-cells and the parabolic reflector is realized by metalized via holes mimicking the role PEC boundary condition.
While the periodic vias with a period of smaller than twice the length of the vias diameter are utilized as the PEC reflector, the bandgap prevents the incoming waves to penetrate from the vias in microwave regime \cite{ke_wu}.
Since the wave is coupled to the surface, a PEC  reflecting boundary with finite height suffices the desired reflection conditions. However, the ideal results may be obtained when the reflector height is infinitely extended. Considering the practical conditions, the reflector height is chosen $1.6 mm$
The radiation pattern results for the simulation and measurement  over a frequency range of 13-18 GHz are shown in Fig.\ref{Results}. The simulation and measurement results have proper agreement indicating the wide-band elevation angle scanning from $33^\circ$ to $69^\circ$. This enables a good scanning application in microwave radar and communication systems instead large phased-array antennas.  
\section{\label{sec:level4}Conclusion}

 In this paper, we explore the holography theorem for leaky wave metasurfaces using surface wave reflectors. Leaky wave holographic metasurfaces can be implemented for beamforming in a specific direction.
By placing a surface wave reflector on a holographic surface, the reference wave modification is necessary. Therefore, the holographic surface impedance distribution is modified depending on the reflector geometry. Full-wave calculation of the surface impedance in presence of the reflector can overcome the restriction to closed-form 2-D scattered fields known for certain reflector geometries and any arbitrary-shaped reflector profile can be chosen to be place on a radiative hologram. Also, full -wave calculation has priority to any approximation of the surface impedance.

 The conventional center-fed holographic antennas are unable of beamforming at the object wave direction when the surface impedance is realized with high accuracy and resolution. It is shown analytically that there is a null in radiation pattern of  the high resolution conventional holograms at the design frequency. Apart from design frequency, the destructive effect non-forward surface leaky wave or  the Rabbit's ear phenomenon causes restricted frequency bandwidth. The utilization of surface wave reflector can remove the null from the radiation pattern by changing the symmetrical distribution of reference wave.
  Parabolic surface wave reflector can redirect the non-forward surface wave and provide considerable operating bandwidth. On the other hand, due to the surface wave confinement resulted from reflector embedding on the metasurface, the diffraction effect of 1-D leaky wave structures are suppressed and better side-lobe level is obtained. In addition, in some application due to the lack of space or other reasons the presence of a reflector on the radiative surface is inevitable. The antenna using the parabolic reflector is enabled for frequency scanning which is confirmed by simulation and measurment. The directivity at different frequencies within the operating bandwidth of the proposed antenna is highly improved compared to the conventional hologram.

\bibliographystyle{unsrt}
\bibliography{Ref.bib}

\end{document}